# The turbulent dynamics of Jupiter's and Saturn's weather layers: order out of chaos?

Peter L Read[*,1], Roland M B Young[1,2], Daniel Kennedy[1,3]

1. Department of Physics, University of Oxford, Clarendon Laboratory, Parks Road, Oxford, OX1 3PU, UK
2. Department of Physics & National Space Science and Technology Center, UAE University, Al Ain, United Arab Emirates
3. Now at Max-Planck-Institut für Plasmaphysik, Greifswald, Germany.

* Corresponding author (peter.read@physics.ox.ac.uk)

## ABSTRACT

The weather layers of the gas giant planets, Jupiter and Saturn, comprise the shallow atmospheric layers that are influenced energetically by a combination of incoming solar radiation and localised latent heating of condensates, as well as by upwelling heat from their planetary interiors. They are also the most accessible regions of those planets to direct observations. Recent analyses in Oxford of cloud-tracked winds on Jupiter have demonstrated that kinetic energy is injected into the weather layer at scales comparable to the Rossby radius of deformation and cascades both upscale, mostly into the extra-tropical zonal jets, and downscale to the smallest resolvable scales in Cassini images. The large-scale flow on both Jupiter and Saturn appears to equilibrate towards a state which is close to marginal instability according to Arnol'd's 2$^{nd}$ stability theorem. This scenario is largely reproduced in a hierarchy of numerical models of giant planet weather layers, including relatively realistic models which seek to predict thermal and dynamical structures using a full set of parameterisations of radiative transfer, interior heat sources and even moist convection. Such models include the Jason GCM, developed in Oxford, which also represents the formation of (energetically passive) clouds of $NH_3$, $NH_4SH$ and $H_2O$ condensates and the transport of condensable tracers. Recent results show some promise in comparison with observations from the Cassini and Juno missions, but some observed features (such as Jupiter's Great Red Spot and other compact ovals) are not yet captured spontaneously by any weather layer model. We review recent work in this vein and discuss a number of open questions for future study.

1. Introduction

The visible atmospheres of Jupiter and Saturn exhibit a bewildering array of features over a huge range of horizontal scales. Both planets are covered in ubiquitous layers of clouds, with the upper layers composed mainly of ammonia ice but with other substances (including water) present at deeper levels (e.g. Irwin 2009). On the largest scales, such clouds are organized into broad bands with meridional widths on the order of $10^4$ km, almost (but not quite) symmetrically disposed about the equator and with boundaries closely aligned with lines of latitude. Such cloud bands are most clearly seen on Jupiter, but also occur in more muted form on Saturn beneath a tenuous layer of upper tropospheric haze. The cloud bands are also now well established to occur in association with an intense and persistent pattern of eastward and westward zonal jet streams, with wind speeds ranging from a few tens of m s$^{-1}$ up to nearly 400 m s$^{-1}$ in Saturn's equatorial jet (e.g. see Ingersoll et al. 2004; Del Genio et al. 2009).

Within this pattern of zonal jets are found a variety of waves and vortices on scales ranging from a few hundred km up to that of the cloud bands at around $10^4$ km. Such waves and vortices are remarkable both for their diversity and also for their persistence and longevity. The most famous of these features is Jupiter's Great Red Spot (GRS), a feature which may be traceable back to a large spot discovered in the 17$^{th}$ century. But more recent observations have identified many other features that resemble the GRS on both Jupiter and Saturn, though on a somewhat smaller scale. The GRS is a giant anticyclone, but persistent cyclonic vortices have also been observed though tend to have a less compact and circular form (at least for those found at low-mid latitudes) than the corresponding anticyclones. Coherent wave trains are also seen from time to time, though few are as persistent and apparently stable and regular as Saturn's North Polar Hexagon. Discovered in images from the Voyager spacecraft in the late 1980s (Godfrey 1988), this feature is characterised by a regular wavenumber m=6 structure around a latitude of 76$^o$ N and has been observed most recently by the Cassini spacecraft to persist coherently up to the end of the mission in 2017 and has since continued to be observable from Earth-based telescopes (Hueso et al. 2020).

Most recently, the Juno mission has significantly extended the observational record to include close orbital passes over both poles of Jupiter, where a whole new set of phenomena have been discovered. At high latitudes, Jupiter's cloud bands and zonal jets are much less evident and the dominant features close to the poles are regular arrays of almost circular, cyclonic vortices surrounding a single cyclone sitting close to the pole itself (Adriani et al. 2018). Infrared images (Adriani et al. 2018) reveal a wealth of small-scale structure associated with such vortices, but most remarkable is the extent to which the 8-fold or 5-fold cyclone arrays have remained stable and persistent over timescales of at least 1-2 years. Other novel findings from Juno have included the discovery that lightning, most likely associated with intense moist convective storms involving latent heat release from condensation of water vapour, is most common at high latitudes in both the northern and southern hemispheres of the planet. Prior to Juno, the only available information on Jovian lightning had noted a prevalence of such storms in the cyclonic cloud "belts" compared to the anticyclonic "zones" (Little et al. 1999), but could not detect larger scale trends with latitude.

Of particular interest from both the Juno and Cassini missions has been the new insights and constraints provided on the vertical structure of the weather layers of Jupiter and Saturn. Until recently, the "weather layer" has been commonly assumed to include the region of the atmosphere influenced by a combination of solar and thermal radiation, upwelling heat from the deep interior and moist convection, typically spanning a height range down to a pressure level of around 10 bars on both planets. However, both the Juno and Cassini missions have included very close passes in high inclination orbits (in Cassini's case shortly before the end of the mission when it was deliberately targeted to enter Saturn's atmosphere) which allowed relatively high order harmonics of each planet's gravity field to be determined. From a comparison with predictions from simple models of the mass distribution associated with pressure variations in balance with hypothetical patterns of eastward and westward zonal winds, estimates have been obtained of the likely depth scale to which the observed cloud level zonal winds could penetrate into the deep interiors of each planet (Kaspi et al. 2018;

Galanti et al. 2019. The values obtained for the depth scale, ~3000 km for Jupiter and ~9000 km for Saturn, are much deeper than had been assumed for the weather layers of either planet, but are consistent with expectations for the depth of their electrically neutral molecular envelopes (Liu & Stephenson 2008; Cao et al. 2017), although uncertainties may still be quite large (e.g. see Kong et al. 2018). Juno has also deployed a relatively long wavelength microwave mapping sensor which has, among other results, shown the perturbations in temperature and $NH_3$ concentration within the GRS to penetrate to depths greater than 300 km (Li et al. 2018). This instrument has also discovered a previously unsuspected, compact and persistent "plume" of $NH_3$ vapour upwelling at the equator, with relatively low concentrations at higher latitudes.

There is now, therefore, a very considerable range of observations from both Jupiter and Saturn available to guide and constrain models of the structure and dynamics of their weather layers, and to stimulate and challenge their further development. In this article, we review the current state of knowledge and model development for these systems, focusing particularly on questions such as:

i. What process(es) primarily drive and energize the cloud-level meteorology on Jupiter and Saturn?
ii. Why does the cloud-level circulation self-organize into patterns of zonal cloud bands and eastward and westward jet streams?
iii. What factors determine the scales and intensities of these zonal jets?
iv. What processes sustain the most energetic waves and vortices in the weather layers of these planets?
v. What are the implications for the transport of heat, momentum and chemical tracers?

And in particular

vi. What aspects of the observable flows can be accounted for solely by the dynamics of a shallow weather layer?

In this article we consider each of these questions, with a particular focus on the energetics of the circulation and the extent to which many of the observations can be captured in a numerical model that is limited to the upper troposphere and stratosphere of a Jupiter-like gas giant planet. Section 2 considers first the strong zonal mean circulation and current ideas for why it takes the form observed. Section 3 discusses some of the implications for the apparent stability or instability of the zonal jets and the distribution of clouds and other tracers. Section 4 reviews the recent development of relatively complex, 3D time-dependent numerical models of jovian weather layers, with examples of some recent successes but also some of their limitations. We conclude in Section 5 with a brief outlook for future research in this area.

2. **Energetics of eddies and zonal jets**

The maintenance of the zonal jets on Jupiter by the action of atmospheric waves and eddies has been established since the time of the Voyager encounters with these planets in the early 1980s (Beebe et al. 1980; Ingersoll et al. 1981). Velocity measurements from manually tracking cloud motions in Voyager ISS images allowed the estimation of zonally averaged horizontal eddy momentum fluxes ($\overline{u'v'}$) and the shear structure of the zonal mean zonal flow ($d\bar{u}/dy$), from which the rate of transfer of kinetic energy into the mean zonal flow could be computed. The measured rate of transfer (typically 1 - 3.5 x $10^{-4}$ W $kg^{-1}$) was initially greeted with some scepticism (e.g. Sromovsky et al. 1982), on the grounds that the manual method of velocity tracking used by Ingersoll et al. (1981) could inadvertently introduce biases in the calculation. This was important, not least because were such a large apparent transfer rate to extend significantly from Jupiter's cloud tops into the deeper atmosphere (e.g. down to pressure levels of ~3 bars) it would imply that more than 10% of the incoming thermal energy from solar and internal heating was being converted into zonal kinetic energy.

More recent work has employed more advanced, automated image correlation techniques to measure velocity fields for both Jupiter (Salyk et al. 2006; Choi & 2011; Galperin et al. 2014) and Saturn (Del Genio & Barbara 2012) using ISS images from the Cassini spacecraft. These have enabled estimates of the eddy-zonal flow kinetic energy transfers at the cloud tops which should be relatively free of some of the observational biases of the manual tracking work. Nevertheless, the results have broadly confirmed the earlier estimates for Jupiter, with an overall conversion rate of around $10^{-4}$ W $kg^{-1}$ at latitudes up to 50º N and S. For Saturn, estimates seem closer to around half of this value (Del Genio & Barbara 2012). The question remains, however, as to how deep this conversion might extend into the abyssal layers on both planets, and hence what is the overall thermodynamic efficiency of the conversion of thermal to zonally averaged mechanical energy.

Resolution of this question (and the corresponding question as to the rate of production of KE from thermally-generated potential energy), however, would require global measurements of velocity and temperature structure that extend much deeper into both atmospheres than has been feasible so far. However, it seems clear that at least the extra-tropical zonal jets on Jupiter and Saturn owe their origin largely to the nonlinear rectification of momentum transfer by eddies. But this raises other related questions such as:
- Do eddies of all scales and locations act in this way to drive zonal flows?
- What is the primary source of the eddies found at the cloud tops?
- What factors set the dominant scales of the zonal jets?
- More generally, how do turbulent cascades of energy and enstrophy operate in gas giant atmospheres?

*2.1 KE spectra*

An important first step in addressing the first of these questions is to determine the kinetic energy spectrum of motions in the atmospheres of Jupiter and Saturn. A number of attempts to estimate spectra for Jupiter's atmosphere have been made in the past (e.g. Mitchell 1982; Choi & Showman 2011), though most of them have focused mainly on estimating these only in the zonal direction. These have provided hints that Jupiter's atmosphere might exhibit evidence for KE spectra resembling the classical Kolmogorov $k^{-5/3}$ spectrum over a certain range of scales, prompting comparisons with expectations from classical theories for 3D or 2D turbulence. However, it is clear from even a superficial inspection of images of Jupiter that the motions in the atmosphere are highly anisotropic with quite different structures in the zonal and meridional directions. A proper analysis should therefore include a consideration of both the zonal and meridional directions, for which the natural decomposition uses the set of spherical harmonics (Choi & Showman 2011; Galperin et al. 2014).

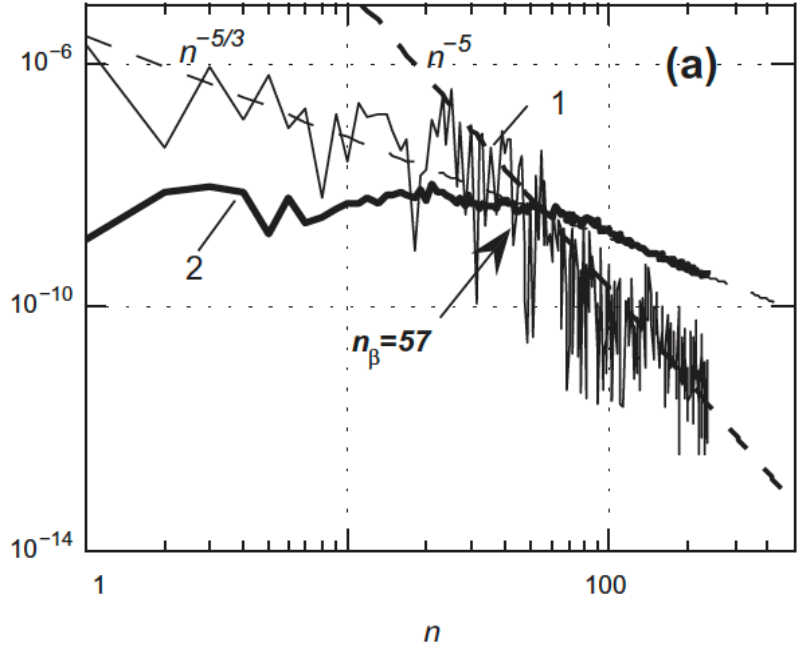

*Figure 1: Kinetic energy spectrum of Jupiter's cloud layer, obtained by Galperin et al. (2014) from Cassini images. Spectra are projected onto spherical harmonics (with total wavenumber index n) and decomposed into the zonal mean flow $E_Z(n)$ (1, thin line) and residual KE $E_R(n)$ (2, thick line), such that total energy $E_T(n) = E_Z(n) + E_R(n)$.*

Figure 1 shows an example of the KE spectrum of Jupiter's atmosphere, derived from automated tracking measurements of cloud motions in Cassini images during the fly-by in December 2000 by Galperin et al. (2014) between latitudes ±50°, in which the KE was projected onto spherical harmonics. This shows KE decomposed into the zonal mean component (for zonal wavenumber $m = 0$, shown with a thin line) and the other ($m \neq 0$) components (thick line). Clear differences are seen between the zonal and residual spectra, with the zonal spectrum roughly following an $n^{-5}$ trend for $n > 20$ and somewhat flatter (nearer $n^{-5/3}$) at larger scales. The residual spectrum, however, is quite different, being almost flat for $n < 50$ and then steepening towards a form close to the Kolmogorov slope $n^{-5/3}$.

This spectrum shows few features in common with that of either an idealised 3D or 2D turbulent fluid. For the latter case, one would expect to see an isotropic spectrum with two segments (e.g. Davidson 2015), one with the Kolmogorov form $n^{-5/3}$ for $n < n_I$ (where $n_I$ represents the scale at which KE is injected, forcing the flow), where the spectrum is dominated by an upscale (spectrally local) cascade of kinetic energy, and the other with a slope of $n^{-3}$ for $n > n_I$, dominated by the downscale transfer of enstrophy (relative or potential). In Jupiter's case, the spectrum is clearly anisotropic, at least at large scales, but for which a Kolmogorov-like segment is only found at small scales in the isotropic component of the residual (eddy) spectrum $E_R$ (and arguably at large scales in the zonal spectrum $E_Z$). At large scales, the KE is heavily dominated by the zonal mean jet circulation, with more than 90% of the total kinetic energy residing in the zonal jets, at least at the cloud tops (Galperin et al. 2014).

The steep $n^{-5}$ character of the zonal spectrum led Sukoriansky et al. (2002) to suggest the relevance of the *zonostrophic* paradigm for geostrophic turbulence, forced at a relatively small scale and dissipated across a range of scales. This represents a generalisation of the classical Kolmogorov-Batchelor-Kraichnan theory to take account of the effects of spherical planetary curvature and the consequent poleward gradients of planetary vorticity and dominance of dispersive Rossby waves at large scales. The consequences for the expected KE spectrum (at least for a predominantly barotropic flow) lead to an anisotropic spectrum with zonal and residual components of the spectrum of the form

$$E_R = C_K \epsilon^{2/3} \left(\frac{n}{R}\right)^{-2/3}, \qquad (1)$$

$$E_Z = C_Z \left(\frac{\Omega}{R}\right)^2 \left(\frac{n}{R}\right)^{-5} \text{ (for n < n}_\beta\text{)}, \qquad (2)$$

where $C_K$ and $C_Z$ are dimensionless constants (O(1) in each case), $R$ is the planetary radius and $\epsilon$ is the upscale kinetic energy transfer rate. The total wavenumber $n_\beta$ is the so-called anisotropy wavenumber, defined in a spherical domain by

$$\frac{n_\beta}{R} = \left[\frac{\Omega^3}{R^3 \epsilon}\right]^{1/5}, \qquad (3)$$

such that the spectrum is anisotropic for n < n$_\beta$ and the two spectra cross over at n = n$_\beta$. This assumes that the turbulent flow is forced at an even smaller scale n$_f$ for which n$_f$ > n$_\beta$.

The zonal spectrum $E_Z$ for both Jupiter and Saturn appear to fit the zonostrophic form in Eq (2) above in both slope and amplitude, given a value for $C_Z \approx 0.5$ (e.g. Sukoriansky et al. 2002; Galperin et al. 2014). For $E_R$ on Jupiter, taking $C_K$ to be approximated by the von Kármán constant $C_K \approx 6$, the observed spectrum for n > n$_\beta$ ~ 57 suggests an estimate for $|\epsilon|$ of around 10$^{-5}$ m$^2$ s$^{-3}$ (Galperin et al. 2014). However, this method of estimating $\epsilon$ does not distinguish between upscale or downscale transfers, leaving open the interpretation of the spectrum as zonostrophic in the sense defined by Sukoriansky et al. (2002) and subsequent work.

2.2 *Spectral transfers and structure functions*

To determine the direction as well as the magnitude of the transfer of energy or enstrophy (squared relative vorticity ½ $\zeta^2$) between different scales within the spectrum, it is necessary to consider higher order covariances of velocities than the quadratic product for the energy spectrum itself. This is commonly analysed within the spectral framework described e.g. by Frisch (1995) or Burgess et al. (2013), for which

$$\frac{\partial \mathcal{E}_n}{\partial t} + \Pi_n = -2\nu \Omega_n + \mathcal{F}_n, \qquad (4)$$

where $\mathcal{E}_n$ represents the cumulative energy contained in scales larger than wavenumber $n$, $\nu$ is the kinematic viscosity and $\Omega_n$ is the cumulative enstrophy over wavenumbers ≤ $n$, so the first term on the left hand side of (4) represents viscous dissipation, and $\mathcal{F}_n$ represents the forcing of kinetic energy at wavenumbers ≤ $n$. The term $\Pi_n$ represents the flux of energy between wavenumbers through wavenumber $n$, due to turbulent interactions between different scales. The sign of $\Pi_n$ is such that $\Pi_n > 0$ implies a direct energy flux towards smaller scales and $\Pi_n < 0$ implies an inverse transfer towards larger scales. $\Pi_n$ is computed (for the rotational, non-divergent component of the flow) from the nonlinear tendency for enstrophy (i.e. the squared vorticity, $1/2\, \zeta_n^{m*} \zeta_n^m$, where $\zeta_n^m$ represents the (*m,n*) component of the vertical component of relative vorticity), as represented for flows decomposed into spherical harmonics, $Y_n^m = P_n(\cos\theta)e^{im\varphi}$ (where $m$ is the zonal wavenumber index and $n$ the total wavenumber index), defined by

$$J_n = -\frac{1}{4}\sum_{m=-n}^{n} \zeta_n^{m*}(\boldsymbol{u}.\boldsymbol{\nabla}\zeta)_n^m + \zeta_n^m(\boldsymbol{u}.\boldsymbol{\nabla}\zeta)_n^{m*}, \qquad (5)$$

where $()^*$ denotes the complex conjugate. This represents the rate of change of enstrophy with total wavenumber $n$ due to quadratic nonlinear interactions with other wavenumber components with zonal index $m$. The corresponding spectral tendency for kinetic energy, $I_n$, can be obtained from the relationship

$$I_n = \frac{a^2}{n(n+1)}J_n, \qquad (6)$$

and the (isotropic) spectral fluxes of energy ($\Pi_n$) and enstrophy ($H_n$) are then obtained from

$$H_{n+1} = -\sum_{\ell=0}^{n} J_\ell; \qquad (7a)$$

$$\Pi_{n+1} = -\sum_{\ell=0}^{n} I_\ell. \qquad (7b)$$

Similar tendencies and fluxes for the divergent, irrotational components of the flow can also be obtained via the tendency of the squared horizontal divergence, also decomposed into spherical harmonics.

Profiles of the spectral fluxes of KE and enstrophy in Jupiter's cloud tops were obtained recently by Young & Read (2017) from the same two-dimensional, horizontal wind vector fields derived from tracking cloud features by Galperin et al. (2014) as mentioned above. Some of the results are illustrated in Figure 2. Fig. 2(a) shows the profile of the spectral enstrophy flux, indicating a positive (forward) cascade at almost all scales down to the resolution of the measurements (around 0.5° in latitude and longitude) with no evidence of an inertial range (i.e. no plateau in $H_n$).

The total kinetic energy transfer is shown in Fig. 2(b) and indicates a much more complicated picture. In particular, it shows a negative (upscale) flux between wavelengths of $2 \times 10^3 - 2 \times 10^4$ km, with a hint of an inertial range plateau around $10^4$ km. The flux of KE changes sign, however, at two scales, (i) at a wavelength $\sim 2$-$3 \times 10^4$ km where energy converges, and (ii) around a wavelength of $2 \times 10^3$ km, from which energy diverges, forming a direct forward cascade at smaller scales. Fig. 2(c) shows the same computation including only eddy-eddy interactions (i.e. with the zonal mean components removed). This shows a qualitatively similar pattern, but with an amplitude 5-10 times smaller. This essentially shows that the strongest interactions are direct transfers between eddies and zonal flows, though the spectrally local, eddy-eddy cascade is of a similar form. The overall picture is consistent with the generation of KE as eddies of a scale close to 2000 km, and an accumulation of KE at scales around 20,000 km, close to the scales of the large-scale zonal bands.

The use of a spherical harmonic decomposition for this analysis assumes that this overall picture for transfers of KE and enstrophy between scales applies across the whole planet, and neglects any local variations in the structure of the energy budget. Young & Read (2017) did, however, also verify the picture derived from spectral fluxes using 3$^{rd}$ order structure functions, which derive the sense and magnitude of energy transfers between scales more directly. The longitudinal structure function, $F_{L3}(r)$, is defined by

$$F_{L3}(r) = \langle \delta u_L^3 \rangle, \qquad (8)$$

where $\langle \delta u_L \rangle = \langle u_L(x + |r|) - u_L(x) \rangle$, angle brackets denote an average over all $x$ and $u_L$ is the velocity component lying along the path joining $x$ and $x + r$. Positive values of $F_{L3}(r)$ indicate a positive correlation between separating particle paths and kinetic energy, and hence a tendency to transfer KE to larger scales (e.g. Frisch 1995; Davidson 2015). For a classical Kolmogorov-Batchelor-Kraichnan (KBK) turbulent spectrum with a $k^{-5/3}$ slope, the third-order structure function is expected to scale as $F_{L3}(r) \sim \Pi r$ (e.g. Davidson 2015). Young & Read's (2017) results show a change in sign of $F_{L3}(r)$ at values of $r$ around 2000-3000 km for both local and global velocity maps towards a nearly linear segment with $r$, confirming the suggested reversal of the KE cascade towards a direct, forward cascade at scales smaller than 2000-3000 km.

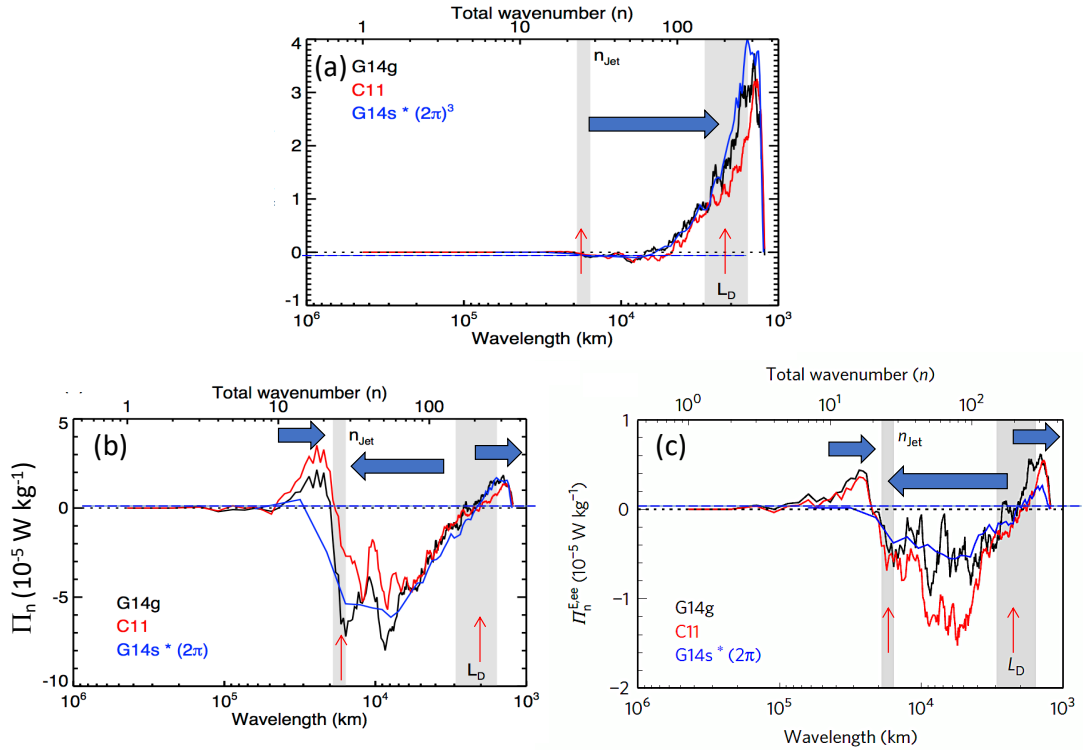

*Figure 2: Profiles of spectral fluxes of enstrophy (squared vorticity) and kinetic energy with total horizontal wavenumber and wavelength at Jupiter's cloud tops, as derived from Cassini ISS images of Jupiter during the 2000 fly-by by Young & Read (2017). (a) enstrophy flux, (b) total (rotational) kinetic energy flux and (c) kinetic energy flux due only to eddy-eddy interactions. The direction of energy or enstrophy transfer is indicated by the arrows.*

The nature of the forward cascade at scales < 2000 km is not well characterised. Such a direct cascade (with a spectrum roughly consistent with the KBK slope of -5/3; see Section 2.1 above) is not consistent with the classical paradigm of 2D or QG turbulence (e.g. Charney 1971; Salmon 1980; Davidson 2015), although a somewhat similar dual cascade pattern is also observed in the Earth's oceans (e.g. Scott & Wang 2005). Internal gravity waves may play an important role if the atmosphere is stably stratified in this region, but observations are insufficient at present to characterise this segment of the cascade with any confidence.

The significance of this injection scale is important with respect to the dominant mechanism that generates the eddies that energise the turbulent cascades. At around 2000-3000 km, this is highly suggestive of the first baroclinic Rossby deformation radius $L_d$. This has been estimated in several previous studies of Jupiter (e.g. Achterberg & Ingersoll 1989; Ingersoll & Kanamori 1995; Ingersoll et al. 2004; Read et al. 2006) to lie around $L_d \sim 2000$ km (to within a factor ~2), consistent with the scale of the KE source found by Young & Read (2017).

Based on Galileo and Cassini observations of Jupiter, Ingersoll et al. (2000) and Gierasch et al. (2000) have suggested the potential significance of moist convection as an energetically important process for generating eddies in Jupiter's atmosphere, estimating that up to 50% of Jupiter's internal heat might be transported upwards through moist convective thunderstorms in the troposphere. However, the typical scale of Jovian thunderstorms is quite a lot smaller than $L_d$, even when aggregated into storm clusters, suggesting that this may not be the principal source of KE in the large-scale turbulent KE cascades measured at Jupiter's cloud tops. More likely would be the possibility that a form of baroclinic instability is responsible for the generation of eddies on scales comparable with $L_d$ (e.g. Vallis 2017), in a similar fashion to the Earth's oceans (e.g. Scott & Wang 2005), although the style of instability may well be rather different from that found more familiarly in terrestrial atmospheres. On Earth, for example, baroclinic instability tends to be strongly enhanced through horizontal thermal gradients at the lower boundary. Such a boundary does not exist on the gas giant planets, however, at least within the upper weather layers. But from the viewpoint of the potential vorticity structure, the tropopause may act as a weak interface, allowing the development of baroclinically unstable modes that are enhanced in amplitude near the top of the tropopause through thermal gradients along the tropopause itself (e.g. Conrath et al. 1981; Read et al. 2020). Many uncertainties remain, however, because the vertical structures of the atmospheres of Jupiter and Saturn are relatively unexplored in observations.

## 2.3 Potential vorticity structure and jet stability

The vorticity structure of the zonal jets on Jupiter and Saturn is of particular interest in relation to their likely stability to breaking up into meanders and/or vortex-like eddies. As we discuss below, it may also hold clues as to why the zonal spectrum, $E_z(n)$, adopts the steep form $\sim n^{-5}$. A long-standing conundrum has been the observation that the latitudinal profile of $\bar{u}(y)$ (where $y$ is the northward coordinate) appears to violate the well known Rayleigh-Kuo necessary (though not sufficient) condition for instability, namely, that $\beta - \partial^2 \bar{u}/\partial y^2 = \partial \zeta_a/\partial y$ changes sign across the domain (where $\beta$ is the northward planetary vorticity gradient due to spherical planetary curvature and $\zeta_a$ is the absolute vorticity). This is clearly seen in Figure 3(b), which shows the latitudinal profiles of northward vorticity gradient (solid lines) and $\beta = 2\Omega \cos\theta /a$ (dashed line). The fact that the solid line crosses the dashed one demonstrates that the Rayleigh-Kuo criterion is violated, principally in association with westward zonal jets.

This tendency for many zonal jets on Jupiter and Saturn was noted in analyses of the early Voyager observations (e.g. Ingersoll et al. 1981), though was explained at the time as indicating that aspects of the vertical structure of the flow (which could not be measured directly) might ameliorate the change of sign of $\partial \zeta_a/\partial y$ (see also Scott & Dunkerton 2017). A more complete stability measure is related to the potential vorticity structure, based on quasi-geostrophic (QG) theory, for which the instability criterion equivalent to the Rayleigh-Kuo condition is the Charney-Stern-Pedlosky (CSP) criterion (e.g. Vallis 2017), namely, that $\partial Q/\partial y$ should change sign within the domain, where

$$Q = \boldsymbol{k}.\nabla \times \boldsymbol{u}_g + f_0 \frac{\partial}{\partial z}\left[\frac{1}{N^2(z)}\frac{\partial \psi_g}{\partial z}\right] + \beta y \qquad (9)$$

is the QG potential vorticity, $\boldsymbol{u}_g$ is the geostrophic velocity and $\psi_g$ the geostrophic stream function. The corresponding profile of $Q$ for Jupiter was derived from a combination of cloud tracked winds and retrieved temperatures from Cassini observations by Read et al. (2006) and is shown in Fig. 3(c). Despite the addition now of realistic (albeit low vertical resolution) thermal information to include the second "stretching" term on the RHS of Eq (9), it is evident that $\partial Q/\partial y$ also changes sign with latitude, suggesting that even the CSP instability criterion is also satisfied. An important caveat for both these criteria is that they are necessary conditions but not sufficient to guarantee instability (e.g. see Vallis 2017). However, the fact remains that the zonal velocity on both gas giant planets robustly satisfies both criteria in repeated measurements.

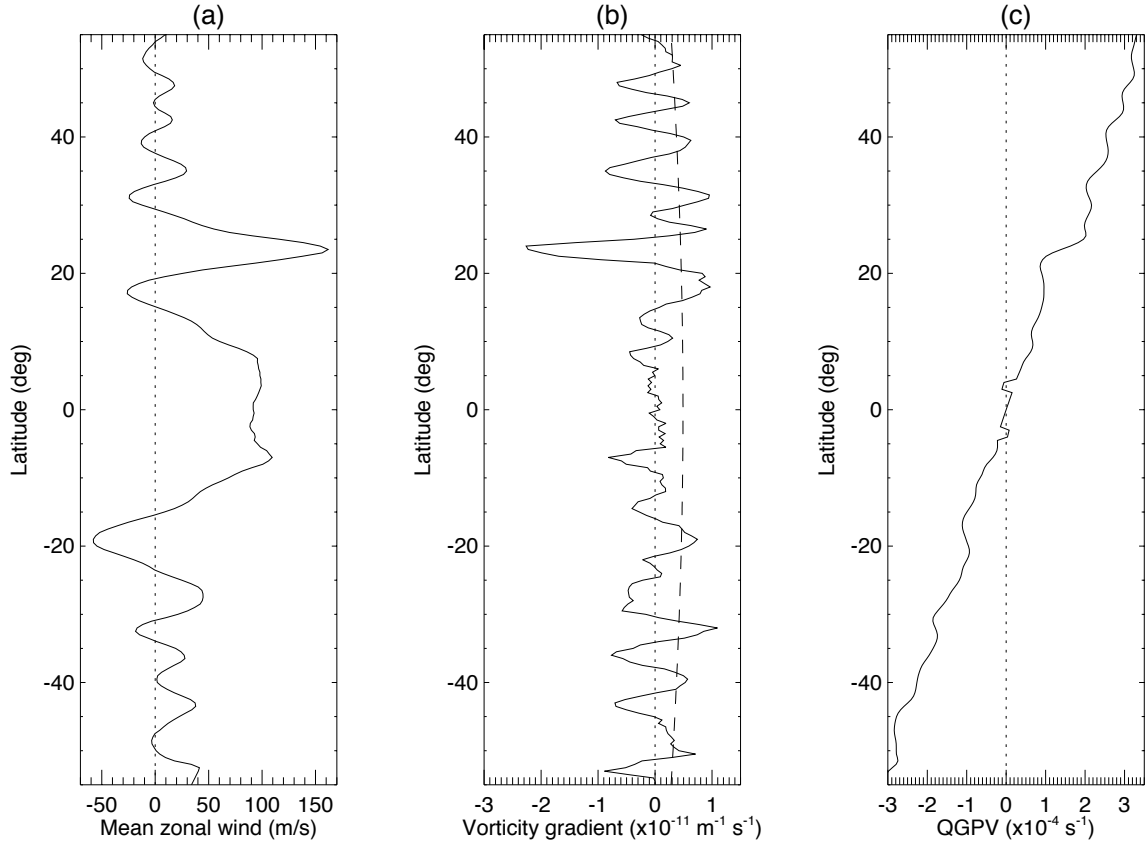

*Figure 3: Observed structure in (a) zonal velocity, (b) northward gradient of relative vorticity and (c) quasi-geostrophic potential vorticity of Jupiter's cloud-level zonal jets, as derived by Read et al. (2006) from Cassini cloud tracking measurements (Porco et al. 2003) and infrared thermal retrievals from the Cassini Composite Infrared Spectrometer (CIRS).*

An alternative interpretation was offered by Stamp & Dowling (1993) and Dowling (1995), who noted that a stronger (nonlinear) stability criterion was provided in the so-called second stability theorem of Arnol'd (Arnol'd 1966) – hereafter "Arnol'd II". Although still only a necessary condition, it does allow for the possibility of flows exhibiting a change of sign of $\partial Q/\partial y$ while remaining marginally stable to non-zonal perturbations. For this situation, the condition for stability reduces to

$$\left(\frac{\bar{u}-\alpha}{\partial Q/\partial y}\right) \geq L_d^2, \tag{10}$$

where $\alpha$ is a constant representing a shift in the frame of reference and $\partial Q/\partial y$ can take either sign. Marginal stability corresponds to the equality condition in Eq (10).

Based on some earlier diagnostic calculations of the vorticity structure of Jupiter's winds (assuming an analogy with rotating shallow water theory), Stamp & Dowling (1993) and Dowling (1995) suggested that both Jupiter's and Saturn's weather layers were close to marginally stable conditions, despite satisfying the Rayleigh-Kuo condition for instability. The essence of this approach relates to the ability of Rossby waves propagating within a local region with a given value of $\bar{u}$ and $\partial Q/\partial y$ to phase lock with a parallel wave train with a different value of $\bar{u}$ and $\partial Q/\partial y$. Marginal instability implies just one value of $\alpha$ that would allow the gravest (deepest) Rossby wave modes to phase-lock, hence implying a unique reference frame in which instability (through interaction between pairs of Rossby wave trains) is possible.

This unique characteristic of a marginally unstable flow was later exploited by Read et al. (2009) to determine the implied reference frame for marginal instability on both Jupiter and Saturn, based on local computations of the zonal mean $Q$ and $\bar{u}$. The results showed that the marginally unstable reference frame on Jupiter was indistinguishable from the System III frame, determined from the rotation of its magnetic field. For Saturn, however, the results indicated a significantly different frame of reference that rotated several minutes faster than the System III frame determined from Voyager observations. With a value of $10^h\ 34^m\ 13\pm20^s$, it implies a shift in apparent zonal velocity on Saturn's equator of around 85 m s$^{-1}$, significantly altering the shape of its zonal velocity profile towards a more symmetrical distribution of $\bar{u}$ between eastward and westward jets, more like Jupiter.

This result was controversial at the time, because Cassini measurements of Saturn's magnetic field had indicated a rather different rotation period to Voyager; a trend that continued throughout the Cassini mission (e.g. Ye et al. 2016), strongly indicating that these measurements were not accurately sensing Saturn's "true" internal rotation period. Coincidentally, however, another method of estimating Saturn's internal rotation rate was also published in 2007 (Anderson & Schubert 2007), using measurements of its oblateness and gravity field, which also inferred an even faster rotation period of $10^h\ 32^m\ 35\pm13^s$. These results have received some further support from measurements of perturbations in Saturn's rings (Mankovitch et al. 2019), which suggest an internal rotation period of $10^h\ 33^m\ 38^s$, though with fairly large uncertainties $\sim\pm2^m$. Nevertheless, it seems fairly clear that, assuming Saturn has a well defined internal rotation period, it must be significantly faster than the System III frame determined by either Voyager or Cassini radio observations. Moreover, the influence of this internal rotation seems to be felt in the dynamical structure of the flow, even in the weather layers of both Jupiter and Saturn.

One further implication of the gas giant weather layers being close to a marginally unstable state relates to the $n^{-5}$ form of the zonal KE spectrum. Planetary rotation allows for the existence of inflection points in the latitudinal profile of $\bar{u}$ while remaining marginally stable under the Rayleigh-Kuo criterion. This may be consistent with a zonal KE spectrum with an index of at least -4 (e.g. Huang et al. 2001), although this makes a number of simplifying assumptions (notably of linearity and a uniform value of $\beta$). Huang et al. (2001) argue that the $n^{-5}$ form of the zonal KE spectrum could represent a saturated flow, in which nonlinearities limit the amplitude of both $\bar{u}$ and its vorticity. The observations, therefore, that both Jupiter and Saturn exhibit an $n^{-5}$ spectrum in their zonal KE of the form in Eq (2) appears to be consistent with the notion that their zonal mean flow structure is defined by a marginally unstable, saturated state.

### 3. Global circulation models of gas giant weather layers

One of the most stringent tests of our understanding of a complex system such as Jupiter's or Saturn's weather layer is the extent to which we can reproduce observed structures and behaviour, and even predict features not yet observed, in a model based on fundamental principles and assumptions, i.e. without overt empirical "tuning". In this regard, the task of developing such models for Jupiter and Saturn is still at a fairly primitive stage, although a rich hierarchy of dynamical models of varying sophistication and complexity have appeared in the past ~40 years.

*3.1 Idealised models*

The earliest numerical circulation models made minimal assumptions concerning the vertical structure of Jupiter's atmosphere, treating it essentially as a single, shallow fluid layer on the surface of a sphere (Williams 1978; Williams 1979; Williams & Yamagata 1984; Williams & Wilson 1988), based on the rotating shallow water equations. Later work led to some account being taken of the likely presence of a deep abyssal layer (assumed to be relatively dynamically passive) underlying a shallow (but dynamically active) weather layer, forming effectively a 2½ layer model (e.g. Ingersoll & Cuong 1981; Dowling & Ingersoll 1988, 1989). This class of model was able to capture certain features and phenomena that were consistent with some observations, such as the formation of patterns of parallel jet streams and the short-term behaviour of some waves and oval vortices, though they were strongly idealised and lacked a representation of a number of realistic physical processes to sustain or dissipate circulation features or their vertical structure. Nevertheless, this class of model has continued to be employed in a number of more recent theoretical studies intended to investigate particular hypotheses for the origin and stability of various features (e.g. Li et al. 2006; Showman 2007; Thomson & McIntyre 2016).

*3.2 Intermediate complexity models*

Because of their computational expense, as well as a lack of detailed observations of the vertical structure of gas giant atmospheres, researchers have been relatively slow to apply full scale 3D numerical models, based on the meteorological Primitive Equations, to this region of gas giant atmospheres. The earliest attempts to apply such 3D models were still quite heavily idealised (Williams 1996, 2002, 2003) though did seek to impose plausible patterns of diabatic forcing and vertical structure to a shallow weather layer. Results did confirm the role of baroclinically unstable eddies in generating patterns of multiple, upper surface-enhanced zonal jets which tended to drift in latitude when heating was confined to upper levels.

Around the same time, Dowling et al. (1998) published the first results from their Explicit Planetary Isentropic Coordinate (EPIC) model, formulated (as the acronym would suggest) using isentropic coordinates in the vertical direction. This was initially applied to a number of idealised problems, mostly in regional rather than global domains, to study the dynamics of artificially initialised vortices resembling Neptune's Great Dark Spot (LeBeau et al. 1998) though has more recently been applied to other features such as Jupiter's equatorial hot spots (Showman & Dowling 2000) and White Ovals (Morales-Juberias et al. 2003), Saturn's ribbon waves and polar hexagon (Morales-Juberias et al. 2011) and other giant planet vortices (e.g. Brueshaber et al. 2019). EPIC has also been extended recently to use hybrid isentropic-pressure coordinates and more realistic parameterisations e.g. for cloud microphysics (Dowling et al. 2006; Palotai et al. 2008).

Other weather layer models have appeared which use a well established dynamical core to solve the dynamical equations but also have used idealised thermal relaxation or mechanical forcing to drive circulations, either in a limited area domain (e.g. Yamazaki et al. 2004; Zuchowski et al. 2009a, based on a dynamical core from the UK Met Office) or globally (e.g. Lian & Showman 2008; Medvedev et al. 2013). These models were able to capture certain features relevant to gas giant meteorology, such as the interaction of waves and zonal flows, though were more diagnostic than predictive. Lian & Showman (2008), for example, showed that forcing localised within a shallow weather layer could nonetheless lead to a much deeper response e.g. in the pattern of zonal winds. Lian & Showman (2010) also introduced a very simple parameterisation to represent the effects of moist convective storms as a random pattern of mass sources and sinks, suggesting the feasibility of driving the zonally banded circulation through nonlinear interactions between moist convection and zonal jets in a rapidly-rotating, stratified spherical atmosphere. But few attempts were made in these studies accurately to match the conditions and forcing processes in the tropospheres or stratospheres of Jupiter or Saturn.

*3.3 More realistic weather layer models*
Most recently, several groups have attempted to develop global circulation models that include a more realistic set of parameterizations that represent forcing and dissipation processes that (arguably!) more closely resemble those found in the upper tropospheres and stratospheres of Jupiter and/or Saturn, and even in some cases including the ice giant planets (Uranus and Neptune).

Lian & Showman (2010), for example, used the MITgcm dynamical core, forced by a combination of thermal relaxation towards a temperature structure inspired by the observed structures of the gas and ice giant atmospheres, trending towards a neutrally stable deep interior (down to a depth ~100 bar), with latent heat release associated with the condensation of a water vapour tracer. The results demonstrated the spontaneous formation of multiple, extra-tropical, eddy-driven zonal jets, together with reasonably strong equatorial jets that reproduced the observed directions for conditions corresponding to either Jupiter, Saturn or Neptune. The equilibrated flow also exhibited a pattern of zonal jets that violated the Charney-Stern stability criterion but in most cases, remained close to a marginally unstable state with respect to Arnol'd's second stability theorem, much as discussed in Section 2.3 above.

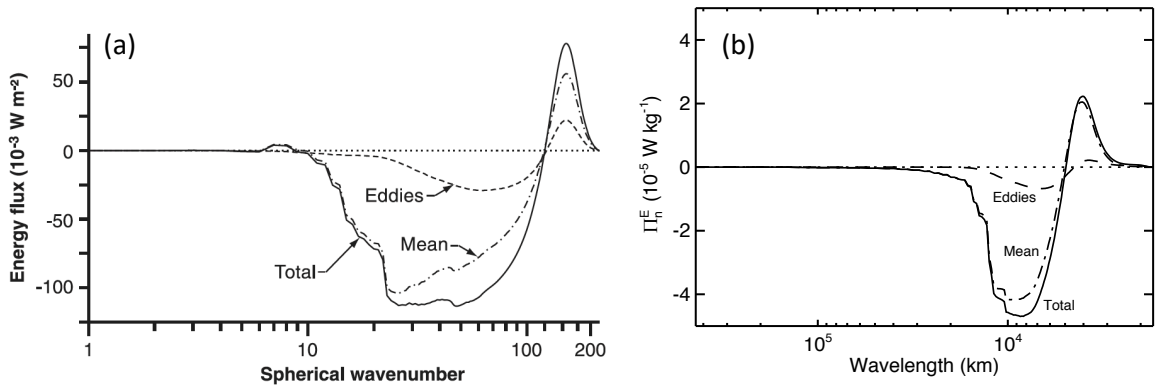

*Figure 4: Spherical harmonic spectral fluxes of KE in model simulations of a Jupiter-like weather layer, obtained by (a) Schneider & Liu (2009) and (b) in the Jason model (Young et al. 2019a, 2020b). Fluxes in (a) are averaged over 3 bar in pressure whereas in (b) this is evaluated (like the observations) at a pressure of 1 bar.*

In a similar vein, Schneider & Liu (2009) used the pseudo-spectral dynamical core from the Flexible Modeling System (FMS) of Princeton's GFDL to obtain simulations of all of the gas and ice giant weather layers (down to a pressure of ~3 bar). In contrast to Lian & Showman (2010), however, they used a semi-gray radiative transfer parameterization with insolation of the observed magnitude (nominally for all four gas and ice giants but for a planet with a fixed Earth-like obliquity) and realistic upward heat fluxes from the interior but no latent heating or moist convection. A pattern of surface drag was also used to emulate magnetohydrodynamic drag at depth to close the energy budget. This model was also able to capture patterns of extra-tropical zonal jets on reasonably realistic spatial scales and amplitudes for each planet, together with equatorial jets of the appropriate sign.

They also computed the exchange of KE between different scales, using the same spectral flux diagnostic as discussed above in Section 2.2. This is illustrated in Figure 4(a), which is taken from Schneider & Liu (2009) and clearly shows a negative (upscale) KE flux between wavenumber indices $10 < n < 100$ turning into a positive (downscale) flux for $n \geq 120$, much as found in the observations of Young & Read (2017). This calculation also shows most of the KE spectral flux residing in the eddy-zonal flow interactions, with only around 20% of the flux in the eddy-eddy components and no obvious evidence of a plateau in spectral flux indicative of an inertial range. Subsequent work (Liu & Schneider 2010, 2015) investigated mechanisms for zonal jet formation in their simulations, highlighting the roles of baroclinic eddies in driving the extra-tropical zonal jets and the relative magnitudes of the interior and solar thermal fluxes in each case in determining the predominant direction (though not necessarily the magnitude) of the equatorial jet, and investigating the scaling of jet strength with model parameters.

Most recently, Spiga et al. (2020) and Cabanes et al. (2020) have developed a new model, based on a new dynamical core on an icosahedral grid (DYNAMICO), that includes a more realistic multi-band (correlated-k) radiative transfer parameterisation which computes heating rates specific to the composition of Saturn's atmosphere (Guerlet et al. 2014). The model also takes account of a realistic heat flux from the deep interior, together with parameterisations for turbulent dissipation and dry convective adjustment, though without moist convection or latent heating. Like the model of Schneider & Liu (2009), the domain only extends to 3 bar pressure with a similar pattern of "MHD drag". Nevertheless, the model was able to capture many features of Saturn's observed thermal structure and meteorology, including its extra-tropical multiple zonal jets and polar vortices. The equatorial jet obtained, however, was not particularly realistic in its magnitude though was at least in the observed direction.

*3.4 The Jason weather layer model*
Young et al. (2019a,b) have recently developed a new GCM for Jupiter's weather layer, designated Jason, which attempts to include a reasonably full range of moderately realistic parameterisations as well as high spatial resolution in its dynamical core. Like the model of Lian & Showman (2008), the dynamical core is based on MITgcm (in its latitude-longitude grid configuration), but follows Schneider & Liu (2009) in utilising a semi-gray radiative transfer scheme that is calibrated to reproduce approximately the observed vertical distribution of solar and infrared radiative fluxes. A uniform internal heat flux is input at the bottom of the domain (at ~18 bar) and the model includes a uniform, weak linear drag at the bottom boundary to emulate "MHD drag" (including in the tropics). The model can also include the effects of latent heat release by condensation of a water vapour tracer and parameterisations of both dry and moist convection (using the heat engine parameterisation of Zuchowski et al. 2009b). As in the earlier work of Zuchowski et al. (2009c), tracers can be advected for water, $NH_3$ and $NH_4SH$ and their condensable cloud aerosols, allowing the model to develop its own layered cloud structures which can then be transported by the dynamics and form features that can resemble observed cloud structures. An example of the modelled distribution of water clouds is illustrated in Figure 5, which shows a snapshot of water ice column density across the planet. Clouds clearly organise into zonal bands with sharp edges, picking out small-scale travelling waves propagating along zonal jets.

Like the other realistic models mentioned above, Jason exhibits the spontaneous formation of a pattern of zonal jets, whose scale and magnitude depends somewhat on which forcing processes and parameterisations are active, as well as on model resolution. Most runs so far have been carried out, as in the other models in Section 3.3, using a resolution of approximately 0.7° x 0.7° in latitude and longitude. This resolution, or higher, is necessary when simulating Jupiter's meteorology in order to resolve adequately the first baroclinic Rossby radius on Jupiter, $L_d$, which on Jupiter is around 1000-3000 km (Achterberg & Ingersoll 1989; Read et al. 2006) and represents the dominant energy-containing scales for baroclinic instabilities on a planet like Jupiter.

As evident in spectral KE flux calculations from the Jason simulations (e.g. see Fig. 4b and Young et al. 2020b), zonal jets arise through predominantly upscale nonlinear transfers of KE from eddies into zonal flows. In the Jason simulations, this takes place between horizontal wavelengths of 20-30,000 km (comparable to the latitudinal spacing of the zonal jets) and ~2-3000 km (comparable to $L_d$). At scales smaller than 2000 km, the KE transfer is direct, towards small scales, much as found in the observations and in the simulations of Schneider & Liu (2009), consistent with a source of KE at scales around 3000 km.

The origin of this KE is indicated from preliminary calculations of the spectral conversion rate of potential energy (PE) to KE, $C_n^E$, which represents the spectrally decomposed covariance of vertical velocity and density (Young et al. 2020b). Preliminary results suggest a significant conversion of APE to KE at scales around 3000 km. If confirmed, this would be consistent with the interpretation that baroclinic eddies on scales comparable with $L_d$ are dominant, not only in generating large-scale structures and zonal jets through upscale nonlinear interactions, but also in energising the KE cascade by conversion from potential energy at small scales in a scenario reminiscent of the QG turbulence paradigm of Salmon (1980). Unlike these classical paradigms, however, the inverse KE cascade to large scales is also accompanied by a forward cascade to even smaller scales, the nature of which is still not well understood but is likely to involve three-dimensional processes related to the stratification.

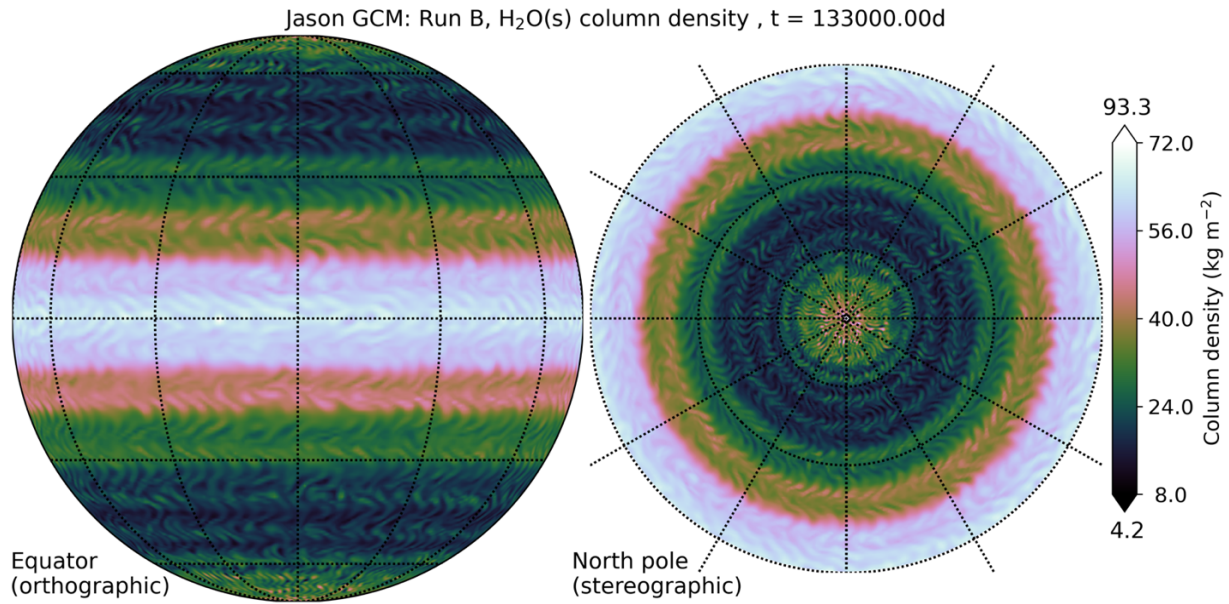

*Figure 5: Visualisation of a snapshot of the field of water cloud ice column density from a simulation of the weather layer circulation of Jupiter at a horizontal resolution of 0.7º x 0.7º in latitude x longitude using the Jason model (Young et al. 2019b). This simulation does not include moist convection.*

Although baroclinic instability may dominate the dynamics, at least at extra-tropical latitudes, moist convection is also active in runs that include the moist convection parameterisation. The distribution of moist convective events, as captured by the parameterisation, is far from uniform, however. Figure 6 shows an example from a Jason simulation in which the Zuchowski et al. (2009b) parameterisation is active (Young & Read 2020a), which shows (a) the net heating/cooling rates due to moist convection as a function of latitude and height, and (b) the vertical distribution of moist convective adjustment during the run. This clearly demonstrates that the most frequent and intense moist convection takes place at high latitudes, with weak heating and cooling close to the water condensation level, around 3-4 bars in this case, and stronger warming at the detrainment levels around 200-600 hPa. This distribution corresponds remarkably well to the observed distribution of lightning, as found recently by Brown et al. (2018) from Juno observations. This tendency to favour moist convection at high latitudes may reflect the stabilising effect of absorbed solar radiation at low latitudes, which tends to suppress deep convection down to levels of a few bars (though not beneath this). One aspect of the observations that is not well captured in the Jason simulations, however, is the significant asymmetry between the northern and southern hemispheres on Jupiter, for which the north seems to have been much more active recently than the south.

The modelled transport and distribution of ammonia vapour is another quantity that is represented in the Jason model (Young et al. 2019b), and can be compared with recent Juno

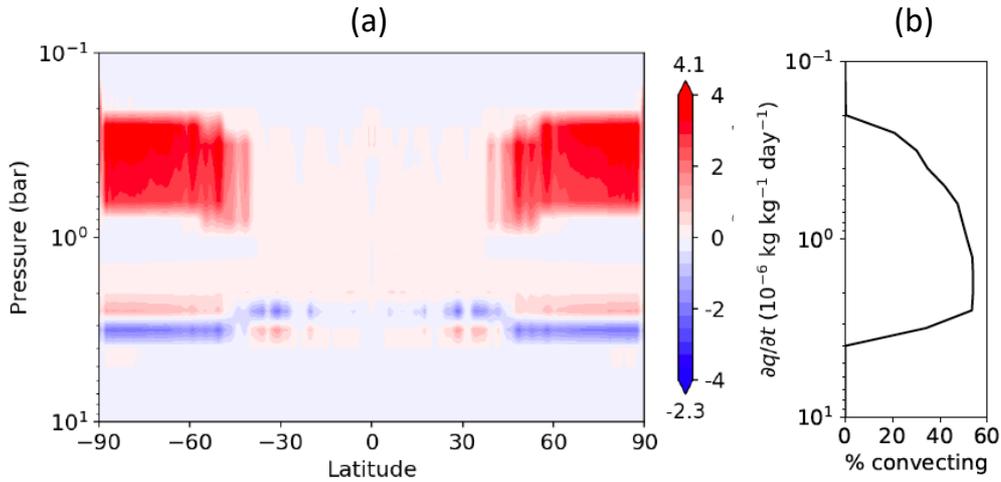

*Figure 6: Distribution of moist convection in a Jason simulation of Jupiter's weather layer at a resolution of 0.7° x 0.7° in latitude x longitude (Young & Read 2020a), showing (a) the net heating and cooling rate as a function of latitude and pressure, and (b) the fraction of convecting gridpoints as a function of pressure, showing the confinement of moist convection between ~4 bars and 200 hPa.*

MicroWave Radiometer (MWR) observations (Li et al. 2018). Figure 7 shows the distribution in latitude and height (pressure) of the $NH_3$ vapour mass mixing ratio 100-200 days after initialisation in a latitudinally uniform state. This simulation clearly captures a strong, ascending plume of ammonia from the dense reservoir in thermochemical equilibrium at depth, capped at around 500 hPa where $NH_3$ vapour tends to condense into clouds. A similar pattern of strongly ascending $NH_3$ vapour was found by Li et al. (2018) in the Juno MWR observations, where a narrow (~5° wide) equatorial ammonia plume was found to rise up from a reservoir at $p > 20$ bar, capped at around 700 hPa. In the Jason simulations, however, the narrow $NH_3$ plume does not persist indefinitely, but tends to spread in latitude as the circulation evolves, equilibrating with a much more diffuse equatorial maximum in $NH_3$ MMR than observed. This almost certainly indicates that, although the model exhibits some realistic tendencies in the organisation of the circulation, there are processes missing (and as yet poorly understood) that either keep the equatorial $NH_3$ confined to a narrow ascending plume or deplete the upper tropospheric $NH_3$ concentration at mid-latitudes (e.g. Guillot et al. 2020).

## 4. Summary and Discussion

The results highlighted in this review paint an increasingly clear picture of how the circulation of Jupiter's and Saturn's weather layers work, although many questions still remain unresolved.

*4.1 Global circulation in weather layers*

Thermodynamic forcing in these weather layers seems to be dominated mainly by a combination of large-scale differential solar heating down to pressures of a few bars and upwelling heat from the planetary interior, balanced by radiative cooling to space. The dynamical response, in the form of baroclinic instabilities and active but complex and anisotropic turbulent cascades of energy and enstrophy, leads to the generation of a pattern of eddy-driven zonal jets with forcing that is probably strongest in the upper troposphere. The response on large horizontal scales, however, seems to penetrate to much deeper levels, as evident in the gravity harmonic results from Juno and Cassini (Kaspi et al. 2018; Galanti et al. 2019).

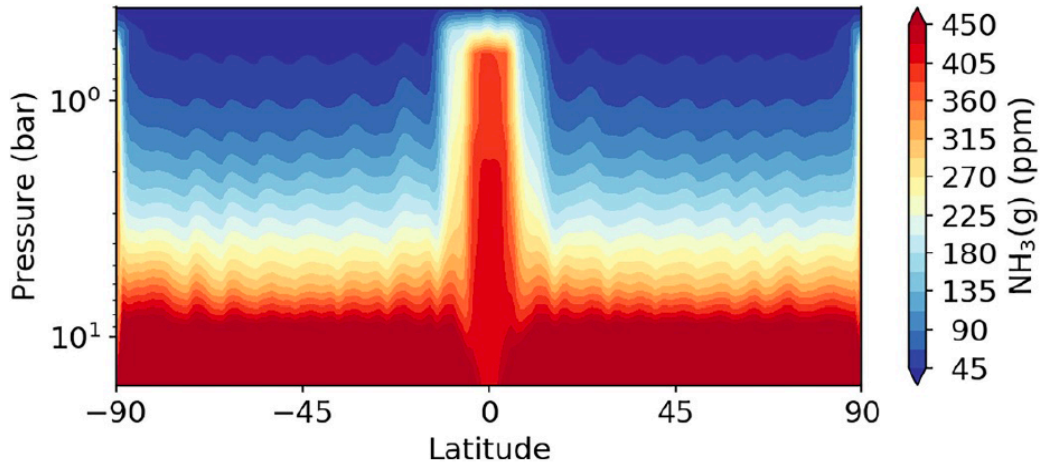

*Figure 7: A map in latitude and height (pressure) of the mass mixing ratio of $NH_3$ vapour obtained 100-200 days after tracers were injected into a Jason model simulation of Jupiter's weather layer, without $H_2O$ latent heating and moist convection (Young et al. 2019b).*

Somewhat remarkably, our most recent models suggest that moist convection, although energetically potentially significant (Gierasch et al. 2000), does not seem to play a dominant role in driving large-scale motions, at least at low-mid latitudes. Young & Read (2020a) find that parameterised moist convection acts to intensify the circulations produced mainly by solar and interior heating towards more realistic intensities but does not seem to result in any new phenomena, at least on global scales (although it may be more important on local or regional scales). This may, of course, be due to the limitations of the parameterisation schemes and limited spatial resolution in the simulations. But the disparity of scales between the typical sizes of individual moist convective storms and the inferred scale of energy injection in the observed turbulent KE cascade (at least for Jupiter; Young & Read 2017, but also likely for Saturn; see Cabanes et al. 2020) would seem to favour baroclinic eddies over convective storms as the principal energising features for large scale flows. However, it will be important in future to establish from observations whether a similar situation prevails on Saturn and the other giant planets.

The main response to such forcing in the weather layer is in the form of the dominant pattern of zonal jets, at least in the extra-tropics, the strength of which may be limited by a tendency for the jet flows to adjust to a state of marginal barotropic instability. Arnol'd's $2^{nd}$ stability criterion would seem to provide a significant constraint (Stamp & Dowling 1993) that has proved effective in pointing towards a particular reference frame in which marginal instability is satisfied for both Jupiter and Saturn (Read et al. 2009a,b). It will be of great interest in future work to explore whether this marginally unstable configuration also occurs on the ice giant planets, Uranus and Neptune, although observations of a sufficient quality to test this hypothesis may need to await a new mission to these remote planets.

*4.2 Atmospheric models*

The most recent generation of numerical weather layer models, using physically-based parameterisations constrained by observations, would seem to offer some significant promise in capturing many, but by no means all, observed features of these atmospheres. The demands on model resources, however, are challenging, requiring quite high horizontal (<1º x 1º in latitude x longitude) and vertical resolution, deep domains (at least to 10-20 bar) and long integration times (>200,000 simulated rotation periods) in order to reach thermal and dynamical equilibrium with realistic heating and cooling rates. Recent results, however, show such approaches to be rewarded by the reproduction of reasonably realistic patterns of extra-tropical zonal jets, cloud bands and certain aspects of meridional tracer transport.

Virtually all weather layer models published to date, however, seem to fail to reproduce a number of major features in observations, at least spontaneously. The large oval vortices on Jupiter and Saturn, such as the Great Red Spot and White Ovals on Jupiter, do not develop spontaneously in these model simulations without careful initialisation, suggesting either some missing physical processes in the model or inadequate spatial resolution. Deep-rooted hot spots might offer one possible explanation, given the appearance of isolated vortical structures in some deep convection models (e.g. see Heimpel et al. 2015). The formation of arrays of polar vortices or symmetric wave patterns, such as Saturn's North Polar Hexagon, have also yet to be demonstrated. The equatorial jets in all four giant planets pose other significant challenges to such models, for which the question remains unresolved as to whether these jets require a deep circulation (e.g. Yano et al. 2005; Heimpel et al. 2005) or can be sustained solely within a shallow weather layer (e.g. Yamazaki et al. 2005).

More generally the connection between shallow weather layers and deep-seated circulations and convection remains poorly understood. Deep models are currently unable to capture the complex structures of the weather layer while shallow weather layer models do not represent a physically realistic lower boundary. The development of coupled deep convection and weather layer models, similar to the coupled atmosphere-ocean models for the Earth's climate, may ultimately be the only way to address this adequately, although this approach also presents major technical challenges.

**Declarations**

**Availability of data and materials**
This is a review article and readers are referred to the original cited references for the majority of material presented. Data from the published Jason model simulations can be accessed at https://doi.org/10.5287/bodleian:PyYbbxpk2. Data relating to the Jupiter cloud winds used in Galperin et al. (2014) and Young & Read (2017) can be accessed from https://ars.els-cdn.com/content/image/1-s2.0-S0019103513003837-mmc2.tar.

**Competing interests**
None to declare.

**Funding**
PLR and RMBY acknowledge support from the UK Science and Technology Research Council during the course of the Oxford-based research reported here under grants ST/K502236/1, ST/K00106X/1 and ST/I001948/1. DK was supported during a summer internship by the UK Met Office Academic Partnership.

**Author contributions**

RMBY led the development of the Jason model, carried out the reported simulations and analysed the results. He also led the analysis of the observational data to compute the spectral fluxes presented in Fig. 2. DK analysed the Jason data for the energy fluxes shown in Fig. 4b. PLR planned, researched and wrote the majority of the paper, with contributions to text and figures from RMBY and DK.

**Acknowledgements**

PLR is grateful to the Asia Oceania Geosciences Society for inspiring this article, which was based on his Distinguished Lecture to the Planetary Sciences section of the 2019 AOGS General Assembly in Singapore. This article is also dedicated to the memory of Prof. Adam Showman, whose untimely passing was announced as this article was being completed.

**References**

Achterberg, R. K. & Ingersoll, A. P. 1989. A Normal-Mode Approach to Jovian Atmospheric Dynamics, J. Atmos. Sci., 46, 2448-2462.

A. Adriani, A. Mura, G. Orton, C. Hansen, F. Altieri, M. L. Moriconi, J. Rogers, G. Eichstädt, T. Momary, A. P. Ingersoll, G. Filacchione, G. Sindoni, F. Tabataba-Vakili, B. M. Dinelli, F. Fabiano, S. J. Bolton, J. E. P. Connerney, S. K. Atreya, J. I. Lunine, F. Tosi, A. Migliorini, D. Grassi, G. Piccioni, R. Noschese, A. Cicchetti, C. Plainaki, A. Olivieri, M. E. O'Neill, D. Turrini, S. Stefani, R. Sordini & M. Amoroso 2018. Clusters of cyclones encircling Jupiter's poles, Nature, 555, 216-219. doi:10.1038/nature25491

Anderson, J. D. & Schubert, G. 2007. Saturn's Gravitational Field, Internal Rotation, and Interior Structure, Science, 317, 1384-1387. Doi:10.1126/science.1144835

Arnol'd, V. I. 1966. On an a priori estimate in the theory of hydrodynamical stability. [In Russian.], Izv. Vyssh. Ucheb. Zaved. Matematika, 54, 3–5.

Beebe, R. .F., A. P. Ingersoll, G, . E. Hunt, J. L. Mitchell & J.-P. Müller 1980. Measurements of wind vectors eddy momentum transports and energy conversions in Jupiter's atmosphere from Voyager 1 images. Geophys. Res. Lett., 7, 1-4.

Brown, S., M. Janssen, V. Adumitroaie, S. Atreya, S. Bolton, S. Gulkis, A. Ingersoll, S. Levin, C. Li, L. Li, J. Lunine, S. Misra, G. Orton, P. Steffes, F. Tabataba-Vakili, I. Kolmašová, M. Imai, O. Santolík, W. Kurth, G. Hospodarsky, D. Gurnett & J. Connerney 2018. Prevalent lightning sferics at 600 megahertz near Jupiter's poles, Nature, 558, 87-90. Doi:10.1038/s41586-018-0156-5

Brueshaber, S. R., Sayanagi, K. M. & Dowling, T. E. 2019. Dynamical regimes of giant planet polar vortices, Icarus, 323, 46-61. Doi: 10.1016/j.icarus.2019.02.001

Burgess, B. H., Erler, A. R. & Shepherd, T. G., 2013. The Troposphere-to-Stratosphere Transition in Kinetic Energy Spectra and Nonlinear Spectral Fluxes as Seen in ECMWF Analyses, J. Atmos. Sci., 70, 669-687.  doi: 10.1175/JAS-D-12-0129.1

Cabanes, S., Spiga, A., & Young, R. M. B., 2020. Global climate modelling of Saturn's atmosphere. Part III: Global statistical picture of zonostrophic turbulence in high-resolution 3D-turbulent simulations, Icarus, 345, 113705. Doi:10.1016/j.icarus.2020.113705

Cao, H. & Stephenson, D. J., 2017. Zonal flow magnetic field interaction in the semi-conducting region of giant planets, Icarus, 296, 59-72. Doi: /10.1016/j.icarus.2017.05.015

Charney, J. G., 1971: Geostrophic turbulence. J. Atmos. Sci., 28, 1087–1095.

Choi, D. S. & Showman, A. P., 2011. Power spectral analysis of Jupiter's clouds and kinetic energy from Cassini. Icarus 216, 597-609

Conrath, B. J., Gierasch, P. J. & Nath, N., 1981. Stability of Zonal Flows on Jupiter, Icarus, 48, 256--282

Davidson, P. A., 2015. Turbulence: An Introduction for Scientists and Engineers 2nd edn, Oxford Univ. Press..

Del Genio, A. D., Achterberg, R. K., Baines, K. H., Flasar, F. M., Ingersoll, A. P., Read, P. L., Sanchez-Lavega, A. & Showman, A. P., 2009. Saturn Atmospheric Structure and Dynamics, in Dougherty, M. et al. (eds). Saturn from Cassini-Huygens, Springer Science & Business Media, pp113-159. Doi:10.1007/978-1-4020-9217-6_6

Del Genio, A.D., Barbara, J.M., 2012. Constraints on Saturn's tropospheric general circulation from Cassini ISS images. Icarus 219, 689–700.

Dowling, T. E., 1993. A relationship between potential vorticity and zonal wind on Jupiter, J. Atmos. Sci., 50, 14-22.

Dowling, T. E. & Ingersoll, A. P., 1988. Potential vorticity and layer thickness variations in the flow around Jupiter's Great Red Spot and White Oval BC. J. Atmos. Sci., 45, 1380-1396.

Dowling, T. E. & Ingersoll, A. P., 1989. Jupiter's Great Red Spot as a shallow water system, J. Atmos. Sci., 46, 3256-3278.

Dowling, T.E., Bradley, M.E., Colón, E., Kramer, J., Lebeau, R.P., Lee, G.C.H., Mattox, T.I., Morales-Juberías, R., Palotai, C.J., Parimi, V.K., Showman, A.P., 2006. The EPIC atmospheric model with an isentropic/terrain-following hybrid vertical coordinate. Icarus 182, 259–273.

Dowling, T.E., Fischer, A.S., Gierasch, P.J., Harrington, J., Lebeau, R.P., Santori, C.M., 1998. The explicit planetary isentropic-coordinate (EPIC) atmospheric model. Icarus 132, 221–238.

Frisch, U., 1995. Turbulence: The Legacy of A. N. Kolmogorov, Cambridge University Press.

Galanti, E., Kaspi, Y., Miguel, Y., Guillot, T., Durante, D., Racioppa, P. & Iess, L., 2019. Saturn's Deep Atmospheric Flows Revealed by the Cassini Grand Finale Gravity Measurements, Geophys. Res. Lett., 46, 616-624. doi:10.1029/2018GL078087

Galperin, B., Young, R.M.B., Sukoriansky, S., Dikovskaya, N., Read, P.L., Lancaster, A.J., Armstrong, D., 2014. Cassini observations reveal a regime of zonostrophic macroturbulence on Jupiter. Icarus 229, 295–320.

Gierasch, P.J., Ingersoll, A.P., Banfield, D., Ewald, S.P., Helfenstein, P., Simon-Miller, A., Vasavada, A., Breneman, H.H., Senske, D.A., Galileo imaging team, 2000. Observation of moist convection in Jupiter's atmosphere. Nature 403, 628–630.

Godfrey, D. A., 1988. A Hexagonal Feature around Saturn's North Pole, Icarus, 76, 335-356

Guerlet, S., Spiga, A., Sylvestre, M., Indurain, M., Fouchet, T., Leconte, J., Millour, E., Wordsworth, R., Capderou, M., Bezard, B., Forget, F., 2014. Global climate modeling of Saturn's atmosphere. Part I: evaluation of the radiative transfer model. Icarus 238, 110–124.

Guillot, T., C. Li, S. J. Bolton, S. T. Brown, A. P. Ingersoll, M. A. Janssen, S. M. Levin, J. I. Lunine, G. S. Orton, P. G. Steffes & D. J. Stevenson, 2020. Storms and the Depletion of Ammonia in Jupiter: II. Explaining the Juno observations, J. Geophys. Res., submitted. Doi: 10.1002/essoar.10502179.1

Heimpel, M., Aurnou, J., Wicht, J., 2005. Simulation of equatorial and high-latitude jets on Jupiter in a deep convection model. Nature 438, 193–196.

Heimpel, M., Gastine, T., Wicht, J., 2015. Simulation of deep-seated zonal jets and shallow vortices in gas giant atmospheres. Nat. Geosci. 9, 19–23. https://doi.org/10.1038/ngeo2601.

Huang, H.-P., Galperin, B. & Sukoriansky, S., 2001. Anisotropic spectra in two-dimensional turbulence on the surface of a rotating sphere. Phys. Fluids 13, 225-240.

Ingersoll, A.P. and Cuong, P.G., Numerical model of long-lived Jovian vortices. J. Atmos. Sci., 1981, 38, 2067–2076.

Ingersoll, A. P., Beebe, R. F., Mitchell, J. L., Garneau, G. W., Yagi, G. M. & J.-P. Müller 1981. Interaction of Eddies and Mean Zonal Flow on Jupiter as Inferred From Voyager 1 and 2 Images, J. Geophys. Res., 86, 8733-8743.

Ingersoll, A.P., Kanamori, H., 1995. Waves from the collisions of comet Shoemaker-Levy 9 with Jupiter. Nature 374, 706–708. https://doi.org/10.1038/374706a0.

Ingersoll, A.P., Dowling, T.E., Gierasch, P.J., Orton, G.S., Read, P.L., Sanchez-Lavega, A., Showman, A.P., Simon-Miller, A.A., Vasavada, A.R., 2004. Dynamics of Jupiter's atmosphere. In: Bagenal, F., Dowling, T.E., McKinnon, W.B. (Eds.), Jupiter: The planet, satellites and magnetosphere. Cambridge University Press, pp. 105–128. https://www.cambridge.org/fr/academic/subjects/physics/computational-scienceand-modelling/jupiter-planet-satellites-and-magnetosphere?format=PB&isbn= 9780521035453

Ingersoll, A.P., Gierasch, P.J., Banfield, D., Vasavada, A.R. and Galileo Imaging Team, Moist convection as an energy source for the large-scale motions in Jupiter's atmosphere. Nature, 2000, 403, 630–632.

Kaspi, Y., Flierl, G.R., 2007. Formation of jets by baroclinic instability on gas planet atmospheres. J. Atmos. Sci. 64, 3177.

Kaspi, Y., Flierl, G.R., Showman, A.P., 2009. The deep wind structure of the giant planets: results from an anelastic general circulation model. Icarus 202, 525–542.

Kaspi, Y., Galanti, E., Hubbard, W.B., Stevenson, D.J., Bolton, S.J., Iess, L., Guillot, T., Bloxham, J., Connerney, J.E.P., Cao, H., Durante, D., Folkner, W.M., Helled, R., Ingersoll, A.P., Levin, S.M., Lunine, J.I., Miguel, Y., Militzer, B., Parisi, M., Wahl, S.M., 2018. Jupiter's atmospheric jet streams extend thousands of kilometres deep. Nature 555, 223–226.

Kong, D., Zhang, K., Schubert, G. & Anderson, J. D., 2018. Origin of Jupiter's cloud-level zonal winds remains a puzzle even after Juno, PNAS, 115, 8499-8504. Doi: 10.1073/pnas.1805927115

LeBeau, R. P. & Dowling, T. E., 1998. EPIC Simulations of Time-Dependent, Three-Dimensional Vortices with Application to Neptune's Great Dark Spot, Icarus, 132, 239-265.

Li, L., Ingersoll, A.P., Huang, X., 2006. Interaction of moist convection with zonal jets on Jupiter and Saturn. Icarus 180, 113–123.

Li, C., A. Ingersoll, M. Janssen, S. Levin, S. Bolton, V. Adumitroaie, M. Allison, J. Arballo, A. Bellotti, S. Brown, S. Ewald, L. Jewell, S. Misra, G. Orton, F. Oyafuso, P. Steffes & R. Williamson, 2017. The distribution of ammonia on Jupiter from a preliminary inversion of Juno microwave radiometer data, Geophys. Res. Lett., 44, 5317–5325, doi:10.1002/2017GL073159

Lian, Y., Showman, A.P., 2008. Deep jets on gas-giant planets. Icarus 194, 597–615.

Lian, Y., Showman, A.P., 2010. Generation of equatorial jets by large-scale latent heating on the giant planets. Icarus 207, 373–393

Little, B., C. D. Anger, A. P. Ingersoll, A. R. Vasavada, D. A. Senske, H. H. Breneman, W. J. Borucki & The Galileo SSI Team, 1999. Galileo Images of Lightning on Jupiter, Icarus, 142, 306–323

Liu, J., Goldreich, P. M. & Stevenson, D. J., 2008. Constraints on deep-seated zonal winds inside Jupiter and Saturn, Icarus, 196, 653-664.

Liu, J., Schneider, T., 2010. Mechanisms of jet formation on the giant planets. J. Atmos. Sci. 67, 3652–3672.

Liu, J., Schneider, T., 2011. Convective generation of equatorial superrotation in planetary atmospheres. J. Atmos. Sci. 68, 2742–2756.

Liu, J., Schneider, T., 2015. Scaling of off-equatorial jets in giant planet atmospheres. J. Atmos. Sci. 72, 389–408.

Mankovitch, C., Marley, M. S., Fortney, J. J., and Movshovitz, N., 2019. Cassini ring seismology as a probe of Saturn's interior. I. rigid rotation, Astrophys. J., 871, 1. https://doi.org/10.3847/1538-4357/aaf798

Medvedev, A.S., Sethunadh, J., Hartogh, P., 2013. From cold to warm gas giants: a three dimensional atmospheric general circulation modeling. Icarus 225, 228–235.

Mitchell, J. L., 1982. The Nature of Large-Scale Turbulence in the Jovian Atmosphere, NASA-CP-169138, 298pp. Jet Propulsion Laboratory, Pasadena, USA.

Morales-Juberías, R., Sánchez-Lavega, A., Dowling, T.E., 2003. EPIC simulations of the merger of Jupiter's White Ovals BE and FA: altitude-dependent behavior. Icarus 166, 63–74.

Morales-Juberías, R., Sayanagi, K.M., Dowling, T.E., Ingersoll, A.P., 2011. Emergence of polar-jet polygons from jet instabilities in a Saturn model. Icarus 211, 1284–1293.

Morales-Juberías, R., Sayanagi, K.M., Simon, A.A., Fletcher, L.N., Cosentino, R.G., 2015. Meandering shallow atmospheric jet as a model of Saturn's north-polar hexagon. Astron. J. Lett. 806, L18.

O'Gorman, P.A., Schneider, T., 2008. Weather-layer dynamics of baroclinic eddies and multiple jets in an idealized general circulation model. J. Atmos. Sci. 65 (2), 524–535.

O'Neill, M.E., Emanuel, K.A., Flierl, G.R., 2015. Polar vortex formation in giant-planet atmospheres due to moist convection. Nat. Geosci. 8, 523–526.

Palotai, C., and T. E. Dowling, 2008. Addition of water and ammonia cloud microphysics to the EPIC model, Icarus, 194, 303–326, doi:1467 10.1016/j.icarus.2007.10.025

Porco, C.C., West, R.A., McEwen, A., Del Genio, A.D., Ingersoll, A.P., Thomas, P., Squyres, S., Dones, L., Murray, C.D., Johnson, T.V., Burns, J.A., Brahic, A., Neukum, G., Veverka, J., Barbara, J.M., Denk, T., Evans, M., Ferrier, J.J., Geissler, P., Helfenstein, P., Roatsch, T., Throop, H., Tiscareno, M., Vasavada, A.R., 2003. Cassini imaging of Jupiter's atmosphere, satellites, and rings. Science 299, 1541–1547. https://doi.org/10.1126/science.1079462.

Read, P.L., Gierasch, P.J., Conrath, B.J., Simon-Miller, A., Fouchet, T., Yamazaki, Y.H., 2006. Mapping potential-vorticity dynamics on Jupiter. I: Zonal-mean circulation from Cassini and Voyager 1 data. Q. J. Roy. Meteor. Soc. 132, 1577–1603. https://doi.org/10.1256/qj.05.34.

Read, P.L., Conrath, B.J., Fletcher, L.N., Gierasch, P.J., Simon-Miller, A.A., Zuchowski, L.C., 2009a. Mapping potential vorticity dynamics on Saturn: zonal mean circulation from Cassini and Voyager data. Planet. Space Sci. 57, 1682–1698.

Read, P.L., Dowling, T.E., Schubert, G., 2009b. Saturn's rotation period from its atmospheric planetary-wave configuration. Nature 460, 608–610.

Read, P., N. Lewis, D. Kennedy, H. Scolan, F. Tabataba-Vakili, Y. Wang, S. Wright & R. Young, 2020. Baroclinic and barotropic instabilities in planetary atmospheres - energetics, equilibration and adjustment, Nonlin. Proc. Geophys., accepted.

Rhines, P.B., 1975. Waves and turbulence on a beta-plane. J. Fluid Mech. 69 (3), 417–443.

Salmon, R., 1980. Baroclinic instability and geostrophic turbulence. Geophys. Astrophys. Fluid Dyn. 15, 167-211.

Salyk, C., Ingersoll, A.P., Lorre, J., Vasavada, A., Del Genio, A.D., 2006. Interaction between eddies and mean flow in Jupiter's atmosphere: analysis of Cassini imaging data. Icarus 185, 430–442.

Schneider, T., Liu, J., 2009. Formation of Jets and Equatorial Superrotation on Jupiter. J. Atmos. Sci. 66 579-+.

Scott, R. B. & Wang, F., 2005. Direct evidence of an oceanic inverse kinetic energy cascade from satellite altimetry. J. Phys. Oceanogr. 35, 1650-1666.

Scott, R. K. & T. J. Dunkerton, 2017. Vertical structure of tropospheric winds on gas giants, Geophys. Res. Lett., 44, 3073–3081, doi:10.1002/2017GL072628.

Showman, A.P., 2007. Numerical Simulations of Forced Shallow-Water Turbulence: Effects of Moist Convection on the Large-Scale Circulation of Jupiter and Saturn. J. Atmos. Sci. 64, 3132.

Showman, A. P. & Dowling, T. E., 2000. Nonlinear Simulations of Jupiter's 5-Micron Hot Spots, Science, 289, 1737-1740. DOI: 10.1126/science.289.5485.1737

Spiga, A., Guerlet, S., Millour, E., Indurain, M., Meurdesoif, Y., Cabanes, S., Dubos, T., Leconte, J., Boissinot, A., Lebonnois, S., Sylvestre, M. & Fouchet, T., 2020. Global climate modeling of Saturn's atmosphere. Part II: Multi-annual high-resolution dynamical simulations, Icarus, 335, 113-377.

Sromovsky, L. A., Revercomb, H. E., Suomi, V. E., Limaye, S. S. & Krauss, R. J., 1982. Jovian winds from Voyager 2. Part II: analysis of eddy transports, J. Atmos. Sci., 39, 1433-1445.

Stamp, A. P. & Dowling, T. E., 1993. Jupiter's winds and Arnol'd's second stability theorem: slowly moving waves and neutral stability, J. Geophys. Res., 98, 18847–18855.

Sukoriansky, S., Galperin, B., Dikovskaya, N., 2002. Universal Spectrum of Two Dimensional Turbulence on a Rotating Sphere and Some Basic Features of Atmospheric Circulation on Giant Planets. Phys. Rev. Lett. 89 (12), 124501.

Thomson, S.I., McIntyre, M.E., 2016. Jupiter's unearthly jets: A new turbulent model exhibiting statistical steadiness without large-scale dissipation. J. Atmos. Sci. 73, 1119–1141. https://doi.org/10.1175/JAS-D-14-0370.1.

Vallis, G. K., 2017. Atmospheric and Oceanic Fluid Dynamics - Fundamentals and Large-Scale Circulation (2nd Edition), Cambridge University Press, Cambridge, UK.

Vasavada, A.R., Showman, A.P., 2005. Jovian atmospheric dynamics: an update after Galileo and Cassini. Rep. Prog. Phys. 68, 1935–1996.

Williams, G.P., Planetary circulations: 1. Barotropic representation of Jovian and terrestrial turbulence. J. Atmos. Sci., 1978, 35, 1339–1426.

Williams, G.P., Planetary circulations: 2. The Jovian quasi–geostrophic regimes. J. Atmos. Sci., 1979, 35, 932–968.

Williams, G. P. & Yamagata, T., 1984. Geostrophic regimes, intermediate solitary vortices and Jovian eddies, J. Atmos. Sci., 41, 453-478.

Williams, G.P., Jovian and comparative atmospheric modeling. Adv. Geophys., 1985, 28A, 381–429.

Williams, G. P., 1996. Planetary vortices and Jupiter's vertical structure, J. Geophys. Res., 102, 9303-9308.

Williams, G. P., 2002. Jovian Dynamics. Part II: The Genesis and Equilibration of Vortex Sets, J. Atmos. Sci., 59, 1356-1370.

Williams, G.P., Jovian dynamics. Part III: Multiple, migrating, and equatorial jets. J. Atmos. Sci., 2003, 60, 1270–1296.

Williams, G.P. and Wilson, R.J., The stability and genesis of Rossby vortices. J. Atmos. Sci., 1988, 45, 207–241.

Yamazaki, Y.H., Skeet, D.R., Read, P.L., 2004. A new general circulation model of Jupiter's atmosphere based on the UKMO Unified Model: Three-dimensional evolution of isolated vortices and zonal jets in mid-latitudes. Planet. Space Sci. 52, 423–445. doi: 10.1016/j.pss.2003.06.006

Yamazaki, Y. H., Skeet, D. R. & Read, P. L., 2005. Hadley circulations and Kelvin wave-driven equatorial jets in the atmospheres of Jupiter and Saturn, Plan. Space Sci., 53, 508–525. DOI: 10.1016/j.pss.2004.03.009

Yano, J.-I., Talagrand, O. & Drossart, P., 2005. Deep two-dimensional turbulence: An idealized model for atmospheric jets of the giant outer planets, Geophysical & Astrophysical Fluid Dynamics, 99, 137-150, DOI: 10.1080/03091920412331336398

Ye, S.-Y., Fischer, G., Kurth, W. S., Menietti, J. D. & Gurnett, D. A., 2016. Rotational modulation of Saturn's radio emissions after equinox. J. Geophys. Res., 121, 11714-11728. Doi: 10.1002/2016JA023281.

Young, R.M.B., Read, P.L., 2017. Forward and inverse kinetic energy cascades in Jupiter's turbulent weather layer. Nat. Phys. 13, 1135–1140.

Young, R.M.B., Read, P.L., Wang, Y., 2019a. Simulating Jupiter's weather layer. Part I: Jet spin-up in a dry atmosphere. Icarus 326, 225–252.

Young, R.M.B., Read, P.L., Wang, Y., 2019b. Simulating Jupiter's weather layer. Part II: Passive ammonia and water cycles. Icarus 326, 253–268.

Young, R. M. B. & Read, P. L., 2020a. Simulating Jupiter's weather layer. Part III: Latent heating and moist convection on a global scale, Icarus, submitted.

Young, R. M. B., Kennedy, D., Read, P. L., 2020b. Simulating Jupiter's weather layer. Part V: The energy cycle. In preparation.

Zuchowski, L.C., Yamazaki, Y.H., Read, P.L., 2009a. Modeling Jupiter's cloud bands and decks: 1. Jet scale meridional circulations. Icarus 200, 548–562. https://doi.org/10.1016/j.icarus.2008.11.024.

Zuchowski, L.C., Read, P.L., Yamazaki, Y.H., Renno, N.O., 2009b. A heat engine based moist convection parametrization for Jupiter. Planet. Space Sci. 57, 1525–1537. https://doi.org/10.1016/j.pss.2009.05.008.

Zuchowski, L.C., Yamazaki, Y.H., Read, P.L., 2009c. Modeling Jupiter's cloud bands and decks. 2. Distribution and motion of condensates. Icarus 200, 563–573. https://doi.org/10.1016/j.icarus.2008.11.015.